\documentclass{article}
\usepackage{latexsym}
\usepackage{epsfig}
\usepackage{amsmath}
\usepackage{amssymb}
\usepackage{amsthm}
\newtheorem{theorem}{Theorem}[section]
\newtheorem{corollary}[theorem]{Corollary}

\newtheorem{prop}[theorem]{Proposition}
\newtheorem{lemma}[theorem]{Lemma}

\theoremstyle{remark}

\theoremstyle{definition}
\newtheorem{definition}[theorem]{Definition}

\DeclareMathOperator\diag{diag}

\begin{document}

\title{Entangled states close to the maximally mixed state}

\author{Roland Hildebrand \thanks{%
LJK, Tour IRMA, 51 rue des
Math\'ematiques, 38400 St.\ Martin d'H\`eres, France ({\tt
roland.hildebrand@imag.fr}).}}

\maketitle

\begin{abstract}
We give improved upper bounds on the radius of the largest ball of separable states of an $m$-qubit system
around the maximally mixed state. The ratio between the upper bound and the best known lower bound
(Hildebrand, quant.ph/0601201) thus shrinks to a constant $c = \sqrt{34/27} \approx 1.122$, as opposed to a
term of order $\sqrt{m\log m}$ for the best upper bound known previously (Aubrun and Szarek,
quant.ph/0503221). We give concrete examples of separable states on the boundary to entanglement which
realize these upper bounds. As a by-product, we compute the radii of the largest balls that fit into the
projective tensor product of four unit balls in $\mathbb R^3$ and in the projective tensor product of an
arbitrary number of unit balls in $\mathbb R^n$ for $n = 2,4,8$.
\end{abstract}

\section{Introduction}

This work deals with balls of separable states around the maximally mixed state for a multi-partite quantum
system consisting of qubits. The best previously known upper bound on the largest radius of such balls was
given in \cite{Szarek0503221} and is of order $\sqrt{m\log m}\cdot 6^{-m/2}$, where $m$ is the number of qubits in
the system. A lower bound proportional to $6^{-m/2}$ was found independently in \cite{Gurvits0409095} and
\cite{Hildebrand0503194}. The proportionality constant $\sqrt{3}$ obtained in these papers was subsequently
improved to $\sqrt{54/17}$ in \cite{Hildebrand0601201}. In this contribution we obtain an upper bound that is
proportional to $6^{-m/2}$ with a constant 2, thus eliminating the factor $\sqrt{m\log m}$ and closing the
gap between the upper and lower bounds up to the multiplicative factor $\sqrt{34/27} \approx 1.122$
(Corollary \ref{upper_bound}). The upper bound in \cite{Szarek0503221} was obtained by computing an upper
bound to the volume radius of the set of separable states and is hence not constructive. However, for the
purpose of quantum computation it is essential to be able to prepare entangled states, and states that are
closer to the maximally mixed state are in general easier to prepare. It is therefore of interest to have
explicit descriptions of entangled states that are close to the maximally mixed state. The upper bound
presented in this contribution is obtained by explicitly constructing mixed states on the boundary to
entanglement that have the distance in question to the maximally mixed state (Corollary
\ref{state_construction}). For a detailed motivation we refer to \cite[Section VI]{Gurvits0409095}.

The main idea underlying the construction is based on using so-called {\it real orthogonal designs} known in
coding theory \cite{TarokhJafarkhaniCalderbank}. The construction involves projective tensor products of unit balls in $\mathbb
R^n$. These objects were introduced in \cite{Szarek0503221}, where a lower bound on the radius of the largest
ball that fits into such a product was given. Namely, the projective tensor product of $m$ unit balls in
$\mathbb R^n$ contains a ball of radius at least $n^{-(m-1)/2}$. As a by-product, we show that this lower
bound is exact for $n = 2,4,8$ and arbitrary $m$ (Theorem \ref{exact_ball}). This is due to the existence in
these dimensions of real orthogonal designs with {\it rate} 1. Further we describe those points on the
boundary of the projective tensor product of 4 unit balls in $\mathbb R^3$ which are closest to the origin
(Corollary \ref{PTP4formD}), and hence obtain also the exact value of the radius of the largest ball in this
product.

\smallskip

The paper is organized as follows. In the next section we provide necessary definitions and summarize known
results which are needed further on. In Section 3 we construct separable mixed states of $m$ qubits that lie
on the boundary to entanglement, but are closer than $2 \cdot 6^{-m/2}$ to the maximally mixed state. In the
appendix we consider the projective tensor product of four unit balls in $\mathbb R^3$ and give a description of those points that lie closest
to the origin. We also provide the matrix of a mixed state of 4 qubits that is produced by our construction.

\section{Definitions and preliminaries}

In this section we recall the definition of separable cones and introduce some related objects.
We also provide some simple properties of these objects.

\medskip

For a vector space $E$ we denote the dual vector space by $E^*$. Let $E_1,\dots,E_n$ be vector spaces. We
denote the tensor product of these spaces by $E_1 \otimes \cdots \otimes E_n$. For the tensor product $E
\otimes \cdots \otimes E$ of $n$ identical spaces we write shorthand $E^{\otimes n}$. The dual space to the
tensor product space $(\mathbb R^n)^{\otimes m}$ is the space of $m$-th order tensors on $\mathbb R^n$, or
otherwise spoken the space of multilinear maps $T: \mathbb R^n \times \cdots \times \mathbb R^n \to \mathbb
R$. The norm $||T||$ of a tensor $T$ will denote the Frobenius norm (the square root of the sum of squares of the components of $T$
in an orthonormal basis).

\subsection{Tensor products of convex bodies}

Let us introduce the following notion from \cite{Szarek0503221}.

{\definition Let $B_1 \subset E_1,\dots,B_n \subset E_n$ be convex bodies residing in finite-dimensional real
vector spaces. Then their {\it projective tensor product} $B = B_1 \otimes \cdots \otimes B_n \subset E_1
\otimes \cdots \otimes E_n$ is defined as the convex hull of the set $\{ x = x_1 \otimes \cdots \otimes x_n
\,|\, x_1 \in B_1,\dots,x_n \in B_n \}$. }

\smallskip

For the projective tensor product $B_1 \otimes \cdots \otimes B_1$ of $n$ identical bodies we write shorthand
$B_1^{\otimes n}$.

{\definition Let $B \subset E$ be a subset of a real vector space. Then the {\it polar} $B^o$ of $B$ is
defined to be the set $\{ x \in E^* \,|\, |\langle x,y \rangle| \leq 1 \}$ for all $y \in B$. }

{\prop \label{dualbody} Let $B_1 \subset E_1,\dots,B_n \subset E_n$ be convex bodies and let $B = B_1 \otimes \cdots \otimes
B_n$ be their projective tensor product. Then the polar $B^o$ is the set of all multilinear maps $T: E_1
\times \cdots \times E_n \to \mathbb R$ such that $|T(x_1,\dots,x_n)| \leq 1$ for all $x_1 \in B_1,\dots,x_n
\in B_n$. }

\begin{proof}
Assume the notations of the proposition.
By definition the scalar product between a multilinear map $T$ and a product element $x_1 \otimes \cdots
\otimes x_n$, $x_1 \in E_1,\dots,x_n \in E_n$ is given by the value of $T(x_1,\dots,x_n)$. But the set
of product elements $\{ x_1 \otimes \cdots \otimes x_n \,|\, x_1 \in B_1,\dots,x_n \in B_n \}$ is exactly
the set of generators of the convex body $B$. The proof concludes with the application of the definition of the polar.
\end{proof}

Let $\partial B$ denote the boundary of $B$.

{\lemma \cite[Lemma 2.10]{Hildebrand0601201} \label{polarrad} Let $B \subset \mathbb R^n$ be a convex,
compact and centrally symmetric body with nonempty interior and define $r_B = \min_{x \in \partial B} |x|$.
Then $(B^o)^o = B$, $B^o$ is convex, compact and centrally symmetric with nonempty interior and $\max_{x \in
B^o} |x| = r_B^{-1}$. Moreover, if $y$ is a unit vector, then $r_B \cdot y \in \partial B$ if and only if
$r_B^{-1} y \in B^o$. \qed }

\subsection{Tensor products of convex cones}

{\definition Let $K_1,\dots,K_n$ be regular convex cones, residing in finite-dimensional real vector spaces
$E_1,\dots,E_n$. Then an element $w \in E_1 \otimes \dots \otimes E_n$ of the tensor product space is called
{\it $K_1\otimes \dots \otimes K_n$-separable} if it can be represented as a finite sum $\sum_{k=1}^N v_1^k
\otimes \dots \otimes v_n^k$ of product elements such that $v_l^k \in K_l$ for all $k = 1,\dots,N$; $l =
1,\dots,n$. The $K_1\otimes \dots \otimes K_n$-separable elements form a regular convex cone, the {\it
$K_1\otimes \dots \otimes K_n$-separable cone}. We denote this cone by $K_1\otimes \dots \otimes K_n$. }

\smallskip

We will speak of the $K_1\otimes \dots \otimes K_n$-separable cone as of the tensor product of the cones
$K_1,\dots,K_n$. For the tensor product $K_1 \otimes \cdots \otimes K_1$ of $n$ identical cones we write
shorthand $K_1^{\otimes n}$.

{\definition Let $E$ be a real vector space equipped with a scalar product $\langle \cdot,\cdot \rangle$ and
let $K \subset E$ be a convex cone. Then the dual cone $K^*$ is defined as the set of elements $y \in E$ such
that $\langle x,y \rangle \geq 0$ for all $x \in K$.}

{\prop \label{dual_cone} Let $K_1 \subset E_1,\dots,K_n \subset E_n$ be convex cones and let $K = K_1 \otimes
\cdots \otimes K_n$ be the corresponding separable cone. Then the dual cone $K^*$ is the set of all
multilinear maps $T: E_1 \times \cdots \times E_n \to \mathbb R$ such that $T(x_1,\dots,x_n) \geq 0$ for all
$x_1 \in K_1,\dots,x_n \in K_n$. }

The proof of this proposition is analogous to the proof of Proposition \ref{dualbody}.

\medskip

We now consider cones generated by convex bodies.

{\definition \label{cone_gen} Let $B \subset {\mathbb R}^{n-1}$ be a convex body. Then the convex cone
\[ K = \left\{ \left. \left( \begin{array}{c} t \\ tx \end{array} \right) \,\right|\, t \geq 0,\ x \in B \right\} \subset \mathbb R^n
\]
is called the cone {\it generated by $B$}. }

\smallskip

In order to work conveniently with cones $K \subset \mathbb R^n$ generated by convex bodies $B \subset \mathbb R^{n-1}$ we introduce the following notations.
The standard basis elements of $\mathbb R^n$ will be denoted by $e_0,\dots,e_{n-1}$, whereas the standard basis elements of $\mathbb R^{n-1}$
will be denoted by $f_1,\dots,f_{n-1}$. Tensors on the space $\mathbb R^n$ will have indices running from $0$ to $n-1$, whereas tensors on the space
$\mathbb R^{n-1}$ will have indices running from $1$ to $n-1$.

Let $B \subset \mathbb R^{n-1}$ be a convex, compact and centrally symmetric body, and let $K \subset \mathbb R^n$ be the cone generated by $B$.
Let $T$ be a tensor of order $m$ on $\mathbb R^n$, satisfying the following conditions:
\[
T_{0\dots0} = 1, \qquad T_{\alpha_1\dots\alpha_m} = 0 \quad \mbox{whenever }\sum_{k=1}^m \alpha_k > 0,\
\prod_{k=1}^m \alpha_k = 0.
\]
Define the $m$-th order tensor $T'$ on $\mathbb R^{n-1}$ by
\[ T'_{\alpha_1\dots\alpha_m} = -T_{\alpha_1\dots\alpha_m} \quad \forall\ 1 \leq \alpha_1,\dots,\alpha_m \leq n-1.
\]

{\lemma Assume above notations. Then the following assertions are equivalent.
\begin{tabbing}
ii) \= \kill
i) \> $T \in (K^{\otimes m})^*$, \\
ii) \> $T' \in (B^{\otimes m})^o$.
\end{tabbing} }

\begin{proof}
i) $\Rightarrow$ ii). Suppose $T \in (K^{\otimes m})^*$ and let $y^1,\dots,y^m \in B$. Let
$y^k_1,\dots,y^k_{n-1}$ be the elements of $y^k$, $k = 1,\dots,m$. Define vectors $x^k \in \mathbb R^n$, $k =
1,\dots,m$ elementwise by $x^k_0 = 1$, $x^k_l = y^k_l$, $l = 1,\dots,n-1$. Then $x^k \in K$ for all $k =
1,\dots,m$ and hence
\begin{eqnarray*}
T(x^1,\dots,x^m) &=& T_{0\dots0} \, x^1_0 \cdot \dots \cdot x^m_0 + \sum_{\alpha_1=1}^{n-1} \cdots \sum_{\alpha_m=1}^{n-1} T_{\alpha_1\dots\alpha_m} x^1_{\alpha_1} \cdot \dots \cdot x^m_{\alpha_m} \\
&=& 1 \, - \sum_{\alpha_1=1}^{n-1} \cdots \sum_{\alpha_m=1}^{n-1} T'_{\alpha_1\dots\alpha_m} y^1_{\alpha_1} \cdot \dots \cdot y^m_{\alpha_m} \geq 0.
\end{eqnarray*}
But then
\[ \langle T', y^1 \otimes \cdots \otimes y^m \rangle = T'(y^1,\dots,y^m) = \sum_{\alpha_1=1}^{n-1} \cdots \sum_{\alpha_m=1}^{n-1} T'_{\alpha_1\dots\alpha_m} y^1_{\alpha_1} \cdot \dots \cdot y^m_{\alpha_m} \leq 1,
\]
and by convexity of $B^{\otimes m}$ we get $\langle T', y \rangle \leq 1$ for all $y \in B^{\otimes m}$. Since $B$ is centrally symmetric, the projective tensor product $B^{\otimes m}$ is also centrally symmetric,
which yields $\langle T', -y \rangle \leq 1$ for all $y \in B^{\otimes m}$. It follows that $T'$ is in the polar of $B^{\otimes m}$.

ii) $\Rightarrow$ i). Suppose $T' \in (B^{\otimes m})^o$ and let $x^1,\dots,x^m \in K$. Let $x^k_0,\dots,x^k_{n-1}$ be the elements of $x^k$, $k = 1,\dots,m$.
If $x^k_0 = 0$ for some index $k$, then by definition of $K$ we have $x^k = 0$ and $T(x^1,\dots,x^m) = 0$. Now suppose that $x^k_0 \not= 0$ for all $k$. By the definition of $K$ we then have $x^k_0 > 0$ for all $k$.
Define vectors $y^1,\dots,y^m \in \mathbb R^{n-1}$ elementwise by $y^k_l = \frac{x^k_l}{x^k_0}$, $k = 1,\dots,m$, $l = 1,\dots,n-1$. Clearly these vectors are elements of $B$. Therefore we have
$T'(y^1,\dots,y^m) \leq 1$ and
\begin{eqnarray*}
T(x^1,\dots,x^m) &=& T_{0\dots0} \, x^1_0 \cdot \dots \cdot x^m_0 + \sum_{\alpha_1=1}^{n-1} \cdots \sum_{\alpha_m=1}^{n-1} T_{\alpha_1\dots\alpha_m} x^1_{\alpha_1} \cdot \dots \cdot x^m_{\alpha_m} \\
&=& x^1_0 \cdot \dots \cdot x^m_0 \left( 1 - \sum_{\alpha_1=1}^{n-1} \cdots \sum_{\alpha_m=1}^{n-1} T'_{\alpha_1\dots\alpha_m} y^1_{\alpha_1} \cdot \dots \cdot y^m_{\alpha_m} \right) \\
&=& x^1_0 \cdot \dots \cdot x^m_0 ( 1 - T'(y^1,\dots,y^m) ) \geq 0.
\end{eqnarray*}
Thus in any case $T(x^1,\dots,x^m) \geq 0$ for every $x^1,\dots,x^m \in K$. Proposition \ref{dual_cone}
completes the proof.
\end{proof}

Let $y = \sum_{\alpha_1=1}^{n-1} \cdots \sum_{\alpha_m=1}^{n-1} y_{\alpha_1\dots\alpha_m} f_{\alpha_1} \otimes \cdots \otimes f_{\alpha_m} \in (\mathbb R^{n-1})^{\otimes m}$. Define an element
$x = e_0 \otimes \cdots \otimes e_0 + \sum_{\alpha_1=1}^{n-1} \cdots \sum_{\alpha_m=1}^{n-1} y_{\alpha_1\dots\alpha_m} e_{\alpha_1} \otimes \cdots \otimes e_{\alpha_m} \in (\mathbb R^n)^{\otimes m}$.

{\lemma \label{ball_cone} Assume above notations. Then the following assertions are equivalent.
\begin{tabbing}
ii) \= \kill
i) \> $x \in K^{\otimes m}$, \\
ii) \> $y \in B^{\otimes m}$.
\end{tabbing} }

\begin{proof}
i) $\Rightarrow$ ii). Suppose $x \in K^{\otimes m}$. Then there exists an integer $N \in \mathbb N$ and vectors $x^{kl} = (x^{kl}_0,\dots,x^{kl}_{n-1})^T \in K$, $k = 1,\dots,m$, $l = 1,\dots,N$ such that $x$ can be represented as
a sum $x = \sum_{l=1}^N x^{1l} \otimes \cdots \otimes x^{ml}$. We can assume without generality that $x^{kl}_0 > 0$ for all $k,l$, otherwise $x^{kl} = 0$ and
the corresponding term in the representation of $x$ is zero and can be omitted.
Define vectors $y^{kl} \in \mathbb R^{n-1}$, $k = 1,\dots,m$, $l = 1,\dots,N$ elementwise by $y^{kl}_{\alpha} = \frac{x^{kl}_{\alpha}}{x^{kl}_0}$, $\alpha = 1,\dots,n-1$. By the definition of $K$ we have
$y^{kl} \in B$ for all $k,l$. Define further elements $y^l \in (\mathbb R^{n-1})^{\otimes m}$, $l = 1,\dots,N$ by $y^l = y^{1l} \otimes \cdots \otimes y^{ml} =
\sum_{\alpha_1=1}^{n-1} \cdots \sum_{\alpha_m=1}^{n-1} y^l_{\alpha_1\dots\alpha_m} f_{\alpha_1} \otimes \cdots \otimes f_{\alpha_m}$. Clearly $y^l \in B^{\otimes m}$ for all $l$.
Let us compute the components of the element $y$. By the definition of $x$ we have
\[ y_{\alpha_1\dots\alpha_m} = \sum_{l=1}^N x^{1l}_{\alpha_1} \cdot \dots \cdot x^{ml}_{\alpha_m} = \sum_{l=1}^N x^{1l}_0 \cdot \dots \cdot x^{ml}_0 y^{1l}_{\alpha_1} \cdot \dots \cdot y^{ml}_{\alpha_m} =
\sum_{l=1}^N x^{1l}_0 \cdot \dots \cdot x^{ml}_0 y^l_{\alpha_1\dots\alpha_m}.
\]
It follows that $y = \sum_{l=1}^N x^{1l}_0 \cdot \dots \cdot x^{ml}_0 y^l$.
Moreover, since the $0\dots0$ element of $x$ equals 1, we have $\sum_{l=1}^N x^{1l}_0 \cdot \dots \cdot x^{ml}_0 = 1$ and $y$ is actually a convex combination of the elements $y^l$.
Thus $y \in B^{\otimes m}$.

ii) $\Rightarrow$ i). Suppose $y \in B^{\otimes m}$. Since $x$ depends affinely on $y$, we can assume without restriction of generality that $y$ can be represented as a product element
$y = y^1 \otimes \cdots \otimes y^m$, where $y^k = (y^k_1,\dots,y^k_{n-1})^T \in B$ for all $k = 1,\dots,m$. Define vectors
$x^{k+} = (1,y^k_1,\dots,y^k_{n-1})^T, x^{k-} = (1,-y^k_1,\dots,-y^k_{n-1})^T \in \mathbb R^n$, $k = 1,\dots,m$.
Clearly $x^{k+} \in K$. Since $B$ is centrally symmetric, we have also $x^{k-} \in K$. For any $m$-tuple $\sigma = (\sigma_1,\dots,\sigma_m) \in \{-1,+1\}^m$ define
$x(\sigma) = x^1(\sigma_1) \otimes \cdots \otimes x^m(\sigma_m)$, where $x^k(\sigma_k)$ is defined by
\[ x^k(\sigma_k) = \left\{ \begin{array}{rcl} x^{k+}, & \quad & \sigma_k = +1, \\ x^{k-}, & \quad & \sigma_k = -1. \end{array} \right.
\]
Then we have $x(\sigma) \in K^{\otimes m}$ for any $\sigma \in \{-1,+1\}^m$. It is not hard to see that $x$ can then be written as
\[ x = \frac{1}{2^{m-1}} \sum_{\sigma \in \{-1,+1\}^m:\,\prod_{k=1}^m \sigma_k = 1} x(\sigma)
\]
and is hence also an element of $K^{\otimes m}$.
\end{proof}

\subsection{Mixed states of multi-qubit systems}

Let ${\cal A}(n)$ be the space of real skew-symmetric $n \times n$ matrices and let $I_n$ be the $n \times n$ identity matrix.
Denote the space of $n \times n$ complex hermitian matrices by ${\cal H}(n)$ and the cone of positive semidefinite matrices in this space by $H_+(n)$. Then we can identify the tensor space
$({\cal H}(n))^{\otimes m}$ with the space ${\cal H}(n^m)$. The tensor product of elements in ${\cal H}(n)$ amounts to the Kronecker product of matrices.

\medskip

The density matrix of a mixed state in an $m$-qubit system is given by a unit trace matrix in $H_+(2^m)$. If
the trace constraint is not satisfied, we speak of an unnormalized density matrix. A state is said to be
separable if its density matrix is contained in the separable cone $H_+(2)^{\otimes m}$. An orthonormal basis
of the space ${\cal H}(2)$ is given by $\{ \frac{\sqrt{2}}{2}\sigma_0, \frac{\sqrt{2}}{2}\sigma_1,
\frac{\sqrt{2}}{2}\sigma_2, \frac{\sqrt{2}}{2}\sigma_3 \}$, where $\sigma_1,\sigma_2,\sigma_3$ are the Pauli
matrices and $\sigma_0 = I_2$. By virtue of this basis we can define an isometry ${\cal I}: \mathbb R^4 \to
{\cal H}(2)$, having the values ${\cal I}(e_k) = \frac{\sqrt{2}}{2}\sigma_k$, $k = 0,1,2,3$ on the basis
vectors of $\mathbb R^4$ and continued by linearity. The preimage ${\cal I}^{-1}[H_+(2)]$ of the positive
semidefinite matrix cone under this isometry is the 4-dimensional Lorentz cone $L_4$, which in turn is
generated by the unit ball $B \subset \mathbb R^3$ according to Definition \ref{cone_gen}.


\section{Mixed states of $m$ qubits}

In this section we construct mixed states of $m$ qubits which lie on the boundary to entanglement and have a
distance of less than $2 \cdot 6^{-m/2}$ to the maximally mixed state.

\medskip

Let ${\cal L}$ be an $n$-dimensional linear subspace of the space $M(N,N)$ of real $N \times N$ matrices
consisting of similarities, i.e.\ for any nonzero matrix $A \in {\cal L}$ there exists a constant $\alpha(A)$
such that $\alpha(A)\cdot A$ is an orthogonal matrix. Let $\{U_1,\dots,U_n\}$ be an orthogonal basis of
${\cal L}$ consisting of orthogonal matrices. Define the linear map ${\cal U}: \mathbb R^n \to {\cal L}$ by
${\cal U}(v) = \sum_{k=1}^n v_kU_k$ for any vector $v = (v_1,\dots,v_n)^T \in \mathbb R^n$. Then ${\cal
U}(v)$ is orthogonal if and only if $||v|| = 1$. Let $O(n)$ be the group of orthogonal $n \times n$ matrices.
The following concept is known from coding theory \cite{TarokhJafarkhaniCalderbank}.

{\definition Let ${\cal L} \subset M(N,N)$ be a linear subspace consisting of similarities and let
$\{U_1,\dots,U_n\} \subset O(N)$ be an orthogonal basis of ${\cal L}$ consisting of orthogonal matrices. If
the elements of the matrices $U_1,\dots,U_n$ are in the set $\{-1,0,+1\}$, then we call the set of matrices
$U_1,\dots,U_n$ a {\sl real orthogonal design}. The ratio $n/N$ is called the {\sl rate} of the design. }

\smallskip

Let us now fix two unit length vectors $w_i,w_f \in \mathbb R^N$ and a number $m \in \mathbb N_+$. Let
$U_1,\dots,U_n$ be as above and consider the multilinear map
\begin{equation} \label{Tm}
T^m: (\mathbb R^n)^m \to \mathbb R, \qquad T^m: (v^1,\dots,v^m) \mapsto w_i^T {\cal U}(v^1) \cdot \dots \cdot
{\cal U}(v^m) w_f.
\end{equation}

{\lemma \label{design_principle} Assume above notations and let $B \subset \mathbb R^n$ be the unit ball. The
tensor $T^m$ defined in (\ref{Tm}) is in $(B^{\otimes m})^o$. If $U_1,\dots,U_n$ form a real orthogonal
design and the elements of the vectors $w_i,w_f$ are in the set $\{-1,0,+1\}$, then $T^m/||T^m||^2 \in
\partial B^{\otimes m}$. If in addition the rate of the design equals 1, then $||T^m||^2 = n^{m-1}$. }

\begin{proof}
Assume the notations of the lemma. Let $v^1,\dots,v^m \in B$ be vectors. Then for every $k = 1,\dots,m$ the
matrix ${\cal U}(v^k)$ has singular values $||v^k||,\dots,||v^k|| \leq 1$. Therefore
\[ ||T^m(v^1,\dots,v^m)|| \leq ||w_i|| \cdot \left( \prod_{k=1}^m ||v^k||\right) \cdot ||w_f||
\leq 1.
\]
By Proposition \ref{dualbody} we then have $T^m \in (B^{\otimes m})^o$, which proves the first assertion of
the lemma.

Let us prove the second one. If $U_1,\dots,U_n$ define a real orthogonal design and the elements of the
vectors $w_i,w_f$ are in the set $\{-1,0,+1\}$, then the components of the tensor $T^m$ can equal only $-1,0$
or $+1$. Let $\eta$ be the number of nonzero components. Then we have $||T^m||^2 = \eta$ and $T^m/||T^m||^2$
has $\eta$ non-zero components, all of which have absolute value $1/\eta$. Hence the tensor $T^m/||T^m||^2$
is an element of the $L_1$ unit ball in $(\mathbb R^n)^{\otimes m}$. But this ball is a subset of $B^{\otimes
m}$, because its extreme points are pure tensor products of unit length vectors. On the other hand, $\langle
T^m, \,c \cdot T^m/||T^m||^2 \rangle = c$ for any number $c \in \mathbb R$. Hence $c \cdot T^m/||T^m||^2
\not\in B^{\otimes m}$ for any $c > 1$. Thus $T^m/||T^m||^2$ must lie on the boundary of $B^{\otimes m}$.

Finally, assume that the rate of the design equals 1, i.e.\ $N = n$. Any of the matrices $U_k$,
$k=1,\dots,n$, has $n$ nonzero entries, otherwise it would not be orthogonal. Therefore out of the $n^3$
entries of the matrices $U_k$ exactly $n^2$ are nonzero. Let us fix two indices $k_1,k_2 \in \{1,\dots,n\}$
and consider the corresponding basis vectors $e_{k_1},e_{k_2} \in \mathbb R^n$. For any unit length vector $v
= (v_1,\dots,v_n)^T \in \mathbb R^n$ we have $||e_{k_1}^T(\sum_{k=1}^n v_kU_k)e_{k_2}|| \leq 1$, because
$\sum_{k=1}^n v_kU_k$ is an orthogonal matrix. Hence there exists at most one index $k_3 \in \{1,\dots,n\}$
such that the $(k_1,k_2)$ element of $U_{k_3}$ equals $-1$ or $+1$. But in order to have $n^2$ nonzero
entries in total, there must exist exactly 1 such index.

Let us now count the nonzero components of the tensor $T^m$. We have
\[ T^m_{\alpha_1\dots\alpha_m} = w_i^T U_{\alpha_1} \cdot \dots \cdot U_{\alpha_m} w_f.
\]
For any $(m-1)$-tuple of indices $(\alpha_2,\dots,\alpha_m)$, the vector $w = U_{\alpha_2} \cdot \dots \cdot
U_{\alpha_m} w_f$ has unit length and entries $-1,0$ or 1. The same holds by assumption for the vector $w_i$.
Hence these vectors are proportional to basis vectors of $\mathbb R^n$, say $e_{k_2}$ and $e_{k_1}$. By the
above, there exists exactly one index $k_3 \in \{1,\dots,n\}$ such that the $(k_1,k_2)$ element of $U_{k_3}$
is nonzero. Therefore $T^m_{\alpha_1\dots\alpha_m} = 0$ for all $\alpha_1 \not= k_3$ and
$|T^m_{\alpha_1\dots\alpha_m}| = 1$ for $\alpha_1 = k_3$. Now note that there are $n^{m-1}$ different
$(m-1)$-tuples of indices $(\alpha_2,\dots,\alpha_m)$. Thus exactly $n^{m-1}$ components of $T^m$ are nonzero
and $||T^m||^2 = n^{m-1}$.
\end{proof}

Recall that for any tensor $T^m \in (B^{\otimes m})^o$, Proposition \ref{dualbody} and Lemma \ref{polarrad}
yield an upper bound on the radius of the largest ball contained in $B^{\otimes m}$, namely $1/||T^m||$. On
the other hand, $n^{-(m-1)/2}$ is a lower bound on this radius \cite[Lemma 4]{Szarek0503221}. From the theory
of Radon-Hurwitz families it is known that real orthogonal designs with rate 1 do exist for the values $n =
1,2,4,8$. We obtain the following result.

{\theorem \label{exact_ball} Let $B \subset \mathbb R^n$ be the unit ball, and let $n$ have one of the values
$1,2,4,8$. Then for any $m \in \mathbb N_+$, the largest ball contained in $B^{\otimes m}$ has radius
$n^{-(m-1)/2}$. \qed }

\medskip

Let us now study the case $n = 4$. A real orthogonal design with rate 1 is defined by the matrices
\begin{equation} \label{design}
U_k = \frac{\partial}{\partial x_k} \left( \begin{array}{cccc} x_4 & -x_3 & x_2 & -x_1 \\ x_3 & x_4 & -x_1 &
-x_2 \\ -x_2 & x_1 & x_4 & -x_3 \\ x_1 & x_2 & x_3 & x_4 \end{array} \right), \qquad k = 1,2,3,4.
\end{equation}

{\lemma \label{sequence} Put the vectors $w_i,w_f$ equal to the basis vector $e_4$ and consider the tensor
$T^m$ defined by (\ref{Tm}) with design (\ref{design}). Then $T^m_{\alpha_1\dots\alpha_m} \not= 0$ if and
only if the indices 1,2,3 appear in the sequence $\alpha_1,\dots,\alpha_m$ either all an even number of
times, or all an odd number of times. }

\begin{proof}
We have the relations $U_4 = I_4$, $U_kU_l = -U_lU_k$ for $1 \leq k \not= l \leq 3$, $U_k^2 = -I_4$ for $k
\not= 4$, and $U_{k_1}U_{k_2} = U_{k_3}$ for any even permutation $(k_1,k_2,k_3) \in S_3$ of $(1,2,3)$. This
leads to the relations
\begin{eqnarray*}
T^m_{\alpha_1\dots \alpha_{k-1} 4 \alpha_{k+1} \dots \alpha_m} &=& T^{m-1}_{\alpha_1\dots \alpha_{k-1} \alpha_{k+1} \dots \alpha_m}, \qquad 1 \leq k \leq m; \\
T^m_{\alpha_1\dots \alpha_{k-2} \alpha \alpha \alpha_{k+1} \dots \alpha_m} &=& -T^{m-2}_{\alpha_1\dots \alpha_{k-2} \alpha_{k+1} \dots \alpha_m}, \qquad 2 \leq k \leq m,\ \alpha \not= 4; \\
T^m_{\alpha_1\dots \alpha_{k-2} \alpha \beta \alpha_{k+1} \dots \alpha_m} &=& -T^m_{\alpha_1\dots \alpha_{k-2} \beta \alpha \alpha_{k+1} \dots \alpha_m}, \qquad 2 \leq k \leq m,\ 1 \leq \alpha \not= \beta \leq 3; \\
T^m_{\alpha_1\dots \alpha_{k-2} \alpha\beta \alpha_{k+1} \dots \alpha_m} &=&
\sigma(\alpha\beta\gamma)T^{m-1}_{\alpha_1\dots \alpha_{k-2} \gamma \alpha_{k+1} \dots \alpha_m}, \qquad 2
\leq k \leq m, \ (\alpha,\beta,\gamma) \in S_3.
\end{eqnarray*}
Here $\sigma(\alpha\beta\gamma)$ is the sign of the permutation $(\alpha,\beta,\gamma)$.

Consider a component $T^m_{\alpha_1\dots\alpha_m}$ of the tensor $T^m$. Above rules allow us to reduce this
component to components of tensors of smaller order. Namely, if the indices 1,2,3 appear in the sequence
$\alpha_1,\dots,\alpha_m$ either all an even number of times, or all an odd number of times, then
$T^m_{\alpha_1\dots\alpha_m}$ eventually reduces to $\pm T^0 = \pm w_i^Tw_f = \pm 1$. If the opposite is the
case, then $T^m_{\alpha_1\dots\alpha_m}$ eventually reduces to $\pm T^1_{\alpha}$, where $\alpha \in
\{1,2,3\}$ is that index whose parity is different from the parities of the two other indices. But $T^1 =
e_4$, and hence $T^m_{\alpha_1\dots\alpha_m} = 0$ in this case.
\end{proof}

Let us now return to the tensor products of balls in $\mathbb R^3$.

{\corollary \label{even_odd} Let $B \subset \mathbb R^3$ be the unit ball. Let $T^m$ be the $m$-th order
tensor on $\mathbb R^4$ from the previous lemma and define an $m$-th order tensor $\tilde T^m$ on $\mathbb
R^3$ by $\tilde T^m_{\alpha_1\dots\alpha_m} = T^m_{\alpha_1\dots\alpha_m}$, $\alpha_1\dots\alpha_m \in
\{1,2,3\}$. Define further an $m$-th order tensor $\hat T^m$ on $\mathbb R^3$ by $\hat
T^m_{\alpha_1\dots\alpha_m} = T^{m+1}_{\alpha_1\dots\alpha_m1}$, $\alpha_1\dots\alpha_m \in \{1,2,3\}$. Then
for any $m \in \mathbb N$ we have $\tilde T^m,\hat T^m \in (B^{\otimes m})^o$; $\tilde T^m/||\tilde
T^m||^2,\hat T^m/||\hat T^m||^2 \in
\partial B^{\otimes m}$. }

\begin{proof}
The tensors $\tilde T^m,\hat T^m$ are defined by the real orthogonal design $(U_1,U_2,U_3)$, where the
matrices $U_k$ are given by (\ref{design}), with $w_i = e_4$, and $w_f = e_4$ for $\tilde T^m$, $w_f = U_1e_4
= -e_1$ for $\hat T^m$. The corollary now follows from Lemma \ref{design_principle}.
\end{proof}

{\lemma \label{Trad} Assume the notations of the previous corollary. Then for even $m$ we have $||\tilde
T^m||^2 = (3^m+3)/4$, and for odd $m$ we have $||\hat T^m||^2 = (3^m+1)/4$. }

\begin{proof}
Let $m$ be even. By the definition of $\tilde T^m$ and by Lemma \ref{sequence} we have $\tilde
T^m_{\alpha_1\dots\alpha_m} \not= 0$ if and only if 1,2,3 all appear in the sequence
$\alpha_1,\dots,\alpha_m$ an even number of times. Hence we get
\begin{eqnarray*}
||\tilde T^m||^2 &=& \sum_{k_1+k_2+k_3 = m/2} \left( \begin{array}{c} m \\ 2k_1\ 2k_2\ 2k_3 \end{array}
\right) = \sum_{k_1 = 0}^{m/2} \left( \begin{array}{c} m \\ 2k_1 \end{array} \right) \sum_{k=0}^{m/2-k_1}
\left( \begin{array}{c} m-2k_1 \\ 2k \end{array} \right) \\ &=& \frac{1}{2} + \sum_{k_1 = 0}^{m/2} \left(
\begin{array}{c} m \\ 2k_1 \end{array} \right) 2^{m-2k_1-1} = \frac{1}{2} \left\{ 1 + \frac{1}{2} \left[
(2+1)^m + (2-1)^m \right] \right\} = \frac{1}{4} \left( 3^m + 3 \right).
\end{eqnarray*}

Let now $m$ be odd. Then $\hat T^m_{\alpha_1\dots\alpha_m} \not= 0$ if and only if 1 appears in the sequence
$\alpha_1,\dots,\alpha_m$ an odd number of times and 2,3 an even number of times. Hence we get
\begin{eqnarray*}
||\hat T^m||^2 &=& \sum_{k_1+k_2+k_3 = (m-1)/2} \left( \begin{array}{c} m \\ 2k_1+1\ \ 2k_2\ \ 2k_3
\end{array} \right) \\ &=& \sum_{k_1 = 0}^{(m-1)/2} \left( \begin{array}{c} m \\ 2k_1+1 \end{array} \right)
\sum_{k=0}^{(m-1)/2-k_1} \left( \begin{array}{c} m-2k_1-1 \\ 2k \end{array} \right) = \frac{1}{2} + \sum_{k_1
= 0}^{(m-1)/2} \left(
\begin{array}{c} m \\ 2k_1+1 \end{array} \right) 2^{m-2k_1-2} \\ &=& \frac{1}{2} \left\{ 1 + \frac{1}{2} \left[
(2+1)^m - (2-1)^m \right] \right\} = \frac{1}{4} \left( 3^m + 1 \right). \qedhere
\end{eqnarray*}
\end{proof}

Define $y = \sum_{\alpha_1,\dots,\alpha_m = 0}^3 y_{\alpha_1\dots\alpha_m} e_{\alpha_1} \otimes \cdots
\otimes e_{\alpha_m} \in (\mathbb R^3)^{\otimes m}$, where the coefficients are given by
\begin{eqnarray} \label{cone_boundary_n}
y_{0\dots0} &=& 1, \nonumber\\
y_{\alpha_1\dots\alpha_m} &=& 0 \quad \mbox{whenever }\sum_{k=1}^m \alpha_k > 0,\ \prod_{k=1}^m \alpha_k = 0, \nonumber\\
y_{\alpha_1\dots\alpha_m} &=& \frac{4}{3^m+3} \tilde T^m_{\alpha_1\dots\alpha_m} \quad \mbox{whenever
}\prod_{k=1}^m \alpha_k > 0,\ m\ \mbox{even}, \nonumber\\
y_{\alpha_1\dots\alpha_m} &=& \frac{4}{3^m+1} \hat T^m_{\alpha_1\dots\alpha_m} \quad \mbox{whenever
}\prod_{k=1}^m \alpha_k > 0,\ m\ \mbox{odd},
\end{eqnarray}
where $\tilde T^m,\hat T^m$ are the tensors defined in Corollary \ref{even_odd}. By combining Corollary
\ref{even_odd}, Lemmata \ref{Trad}, \ref{ball_cone} and using the fact that the cone $L_4$ is generated by
the unit ball $B \subset \mathbb R^3$ we obtain the following result.

{\lemma The element $y \in (\mathbb R^4)^{\otimes m}$ defined in (\ref{cone_boundary_n}) lies on the boundary
of the cone $L_4^{\otimes m}$. \qed }

\smallskip

Recall that there exists an isometry ${\cal I}$ between $\mathbb R^4$ and ${\cal H}(2)$ that takes the
cone $L_4$ to the cone of positive semidefinite matrices $H_+(2)$.

{\corollary \label{state_construction} Let $m \in \mathbb N_+$. The matrix $\rho = \frac{1}{2^{m/2}}{\cal
I}^{\otimes m}(y)$, where $y \in (\mathbb R^4)^{\otimes m}$ is given by (\ref{cone_boundary_n}), describes a
separable mixed $m$-qubit state on the boundary to entanglement. If $m$ is even, then $\rho$ has a distance
of $2^{-m/2}\sqrt{\frac{4}{3^m+3}}$ from the maximally mixed state $2^{-m}\cdot I_{2^m}$. If $m$ is odd, then
$\rho$ has a distance of $2^{-m/2}\sqrt{\frac{4}{3^m+1}}$ from the maximally mixed state. }

\begin{proof}
The first statement follows from the fact that the map $\frac{1}{2^{m/2}}{\cal I}^{\otimes m}$ takes the cone
$L_4^{\otimes m}$ to the cone $H_+(2)^{\otimes m}$.

Let us show the second statement. By definition the tensor $y$ has a distance of $||\tilde T^m||^{-1}$ from
$e_0^{\otimes m}$ for even $m$ and of $||\hat T^m||^{-1}$ for odd $m$. The maximally mixed state is the image
of $e_0^{\otimes m}$ under the map $\frac{1}{2^{m/2}}{\cal I}^{\otimes m}$. This map reduces all distances by
a factor of $2^{m/2}$, because ${\cal I}^{\otimes m}$ is an isometry. Application of Lemma \ref{Trad}
completes the proof.
\end{proof}

{\corollary \label{upper_bound} The radius of the largest separable ball of mixed states around the maximally
mixed state for an $m$-qubit system is bounded above by
\[ r_{upper} = \left\{ \begin{array}{rcl} 2\cdot 6^{-m/2} \sqrt{\frac{1}{1+3^{-m+1}}}, & \quad & m\ \mbox{even}, \\
2\cdot 6^{-m/2} \sqrt{\frac{1}{1+3^{-m}}}, & \quad & m\ \mbox{odd}.
\end{array} \right. \qedhere
\] }

In any case, the radius of the largest separable ball is bounded above by $2 \cdot 6^{-m/2}$.

In \cite{Hildebrand0601201} we derived the following lower bound on this radius:
\[ r_{lower} = {\sqrt{\frac{54}{17}}}\cdot 6^{-m/2}.
\]
The ratio of these bounds then obeys
\[ \frac{r_{upper}}{r_{lower}} < 2{\sqrt{\frac{17}{54}}} = {\sqrt{\frac{34}{27}}} \approx 1.122.
\]
The radius of the largest separable ball around the maximally mixed state is thus determined up to a constant
factor.


\appendix

\section{Projective tensor products of four unit balls}

In this section we study the projective tensor product of four unit balls in $\mathbb R^3$.

\medskip

%

Let $B \subset \mathbb R^3$ be the unit ball. Let us explicitly write out the components of the tensor $\tilde T^4 = T \in (B^{\otimes 4})^o$
defined in Corollary \ref{even_odd}.
\begin{eqnarray} \label{T}
T_{\alpha\alpha\beta\beta} &=& 1 \quad \forall\ \alpha,\beta = 1,2,3; \nonumber\\
T_{\alpha\beta\alpha\beta} &=& -1 \quad \forall\ 1 \leq \alpha \not= \beta \leq 3; \nonumber\\
T_{\alpha\beta\beta\alpha} &=& 1 \quad \forall\ 1 \leq \alpha \not= \beta \leq 3; \nonumber\\
T_{\alpha\beta\gamma\delta} &=& 0 \quad \mbox{for all other index quadruples }(\alpha,\beta,\gamma,\delta).
\end{eqnarray}

By Lemma \ref{Trad} we have $||T|| = \sqrt{21}$.
We shall show that the tensor $T$ is essentially the only element in $(B^{\otimes 4})^o$ which has norm $\sqrt{21}$.

A similar result for $(B^{\otimes 3})^o$ was proven in \cite{Hildebrand0601201}. Namely, Lemma \ref{polarrad} and Theorems 3.8 and 3.9 in \cite{Hildebrand0601201} yield the following result.

{\lemma \label{PTP3form} Let $B \subset {\mathbb R}^3$ be the unit ball. Then $\max_{M' \in (B^{\otimes 3})^o} ||M'|| = \sqrt{7}$. For any $M' \in (B^{\otimes 3})^o$ such that $||M'|| = \sqrt{7}$
there exist orthogonal $3 \times 3$ matrices $U,V,W$ such that $M' = (U \otimes V \otimes W)(M)$, where the tensor $M$ is given by
\begin{eqnarray} \label{M}
M_{111} &=& 1; \nonumber\\
M_{1\alpha\alpha} &=& -1, \quad \alpha = 2,3; \nonumber\\
M_{\alpha1\alpha} &=& -1, \quad \alpha = 2,3; \nonumber\\
M_{\alpha\alpha1} &=& -1, \quad \alpha = 2,3; \nonumber\\
M_{\alpha\beta\gamma} &=& 0, \quad \mbox{for all other index triples}.
\end{eqnarray} }

The lemma implies that the tensor $M$ given by (\ref{M}) lies on the boundary of $(B^{\otimes 3})^o$. The next result describes the face of $B^{\otimes 3}$ which is dual to $M$. This face is generated by the set
$\{ x \otimes y \otimes z \,|\, x,y,z \in B,\, M(x,y,z) = 1 \}$.

{\lemma \label{dualfaceM} Let the tensor $M$ be defined as in (\ref{M}). Then for $x,y,z \in B$ we have
$M(x,y,z) = 1$ if and only if there exist angles $\xi_1,\xi_2,\xi_3,\varphi$ such that $\xi_1+\xi_2+\xi_3 =
0$ and
\begin{equation} \label{xyzform}
x = \left( \begin{array}{c} \cos\xi_1 \\ \cos\varphi\sin\xi_1 \\
\sin\varphi\sin\xi_1 \end{array} \right), \quad y = \left( \begin{array}{c} \cos\xi_2 \\ \cos\varphi\sin\xi_2
\\ \sin\varphi\sin\xi_2 \end{array} \right), \quad z = \left( \begin{array}{c} \cos\xi_3 \\
\cos\varphi\sin\xi_3 \\ \sin\varphi\sin\xi_3
\end{array} \right).
\end{equation} }

\begin{proof}
Assume the conditions of the lemma. The condition $M(x,y,z) = 1$ can be rewritten as
\[ x^T \left( \begin{array}{ccc} z_1 & -z_2 & -z_3 \\ -z_2 & -z_1 & 0 \\ -z_3 & 0 & -z_1 \end{array} \right) y = 1.
\]
Here $z_k$, $k = 1,2,3$ are the elements of $z$. The eigenvalues of the symmetric $3 \times 3$ matrix in the
middle are $\lambda_1 = \sqrt{z_1^2+z_2^2+z_3^2}$, $\lambda_2 = -\lambda_1$ and $\lambda_3 = -z_1$. If $z_2 =
z_3 = 0$, then the corresponding eigenvectors are the basis vectors $f_1,f_2,f_3$. If $z_2^2 + z_3^2 > 0$,
then the normalized eigenvectors are
\[ \left( \begin{array}{c} \frac{-z_1-\lambda}{\sqrt{2\lambda(z_1+\lambda)}} \\ \frac{z_2}{\sqrt{2\lambda(z_1+\lambda)}} \\
\frac{z_3}{\sqrt{2\lambda(z_1+\lambda)}} \end{array} \right), \ \lambda = \lambda_1,\lambda_2; \quad \left(
\begin{array}{c} 0 \\ \frac{z_3}{\sqrt{z_2^2+z_3^2}} \\ -\frac{z_2}{\sqrt{z_2^2+z_3^2}} \end{array} \right),\
\lambda = \lambda_3.
\]
An inspection of this eigenvalue decomposition reveals that the condition $M(x,y,z) = 1$ implies $|x| = |y| =
|z| = 1$ and the last two components of the three vectors $x,y,z$ must all be proportional. Therefore these
vectors have the form (\ref{xyzform}). Insertion of (\ref{xyzform}) in the equation $M(x,y,z) = 1$ yields
\[ \cos\xi_1\cos\xi_2\cos\xi_3 - \cos\xi_1\sin\xi_2\sin\xi_3 - \cos\xi_2\sin\xi_1\sin\xi_3 -
\cos\xi_3\sin\xi_1\sin\xi_2 = \cos(\xi_1+\xi_2+\xi_3) = 1.
\]
This completes the proof.
\end{proof}

{\lemma \label{nullcompl} Let $M' \in (B^{\otimes 3})^o$ be such that $||M'|| = \sqrt{7}$ and $M'(x,y,z) = 0$ whenever $x,y,z
\in B$ and $M(x,y,z) = 1$, where $M$ is the tensor defined by (\ref{M}). Then there exist $\zeta \in {\mathbb
R}$, $\sigma_k,\pi_k \in \{+1,-1\}$, $k = 1,2,3$ such that
\begin{eqnarray} \label{M'cond}
M'_{112} = M'_{121} = M'_{211} = -M'_{222} &=& \cos\zeta, \nonumber\\
M'_{113} = M'_{131} = M'_{311} = -M'_{333} &=& \sin\zeta, \nonumber\\
M'_{111} = M'_{1\alpha\beta} = M'_{\alpha1\beta} = M'_{\alpha\beta1} &=& 0,\ \alpha,\beta = 2,3; \nonumber\\
M'_{322} = \sigma_1\sin\zeta, M'_{232} = \sigma_2\sin\zeta, M'_{223} = \sigma_3\sin\zeta, && \sum_{k=1}^3
\sigma_k = -1; \nonumber\\
M'_{233} = \pi_1\cos\zeta, M'_{323} = \pi_2\cos\zeta, M'_{332} = \pi_3\cos\zeta, && \sum_{k=1}^3 \pi_k = -1.
\end{eqnarray}
}

\begin{proof}
Assume the conditions of the lemma. Let $x_k,y_k,z_k$, $k = 1,2,3$ be the components of the vectors $x,y,z$.

Let us first define $x,y,z$ by putting $\xi_1 = 0$ and $\xi_3 = -\xi_2$ in (\ref{xyzform}). Then $x = f_1$,
$z_1 = y_1$, $z_2 = -y_2$, $z_3 = -y_3$ and $|y| = 1$. By the preceding lemma any triple $(x,y,z)$ satisfying
these relations fulfills the conditions $x,y,z \in B$ and $M(x,y,z) = 1$. Hence we have
\[
M'(x,y,z) = \sum_{\alpha=1}^3 M'_{1\alpha1} y_{\alpha} y_1 - \sum_{\alpha=1}^3 \sum_{\beta=2}^3
M'_{1\alpha\beta} y_{\alpha} y_{\beta} = 0
\]
for any unit length vector $y \in \mathbb R^3$. By homogeneity the condition $|y| = 1$ can be dropped and
above quadratic form in the variables $y_1,y_2,y_3$ must be identically zero. This leads to the conditions
\[ M'_{1\alpha\alpha} = 0,\ M'_{1\alpha1} = M'_{11\alpha} \ \forall \ \alpha; \quad M'_{123} = -M'_{132}.
\]
Similar relations on the components $M'_{\alpha1\beta}$ and $M'_{\alpha\beta1}$, $\alpha,\beta = 1,2,3$ are
obtained by putting $\xi_2 = 0$ or $\xi_3 = 0$ in (\ref{xyzform}) and following the same lines of reasoning.
Combining these results yields the existence of $a,b,c_1,c_2,c_3 \in \mathbb R$ such that
\begin{eqnarray*}
M'_{112} = M'_{121} = M'_{211} &=& a, \\
M'_{113} = M'_{131} = M'_{311} &=& b, \\
M'_{1\alpha\alpha} = M'_{\alpha1\alpha} = M'_{\alpha\alpha1} &=& 0,\ \alpha = 1,2,3; \\
M'_{123} = -M'_{132} &=& c_1, \\
M'_{213} = -M'_{312} &=& c_2, \\
M'_{231} = -M'_{321} &=& c_3.
\end{eqnarray*}
Inserting this and (\ref{xyzform}) in the relation $M'(x,y,z) = 0$ yields
\begin{eqnarray*}
&& (a\cos\varphi + b\sin\varphi)(\cos\xi_1\cos\xi_2\sin\xi_3 + \cos\xi_1\sin\xi_2\cos\xi_3 + \sin\xi_1\cos\xi_2\cos\xi_3) \\
&+& \sin\xi_1\sin\xi_2\sin\xi_3(M'_{222}\cos^3\varphi + (M'_{322} + M'_{232} + M'_{223})\cos^2\varphi\sin\varphi \\
&+& (M'_{233} + M'_{323} + M'_{332})\cos\varphi\sin^2\varphi + M'_{333}\sin^3\varphi) = 0.
\end{eqnarray*}
This relation must be valid for all $\varphi,\xi_1,\xi_2,\xi_3$ such that $\xi_1+\xi_2+\xi_3 = 0$. Using this
last relation, we obtain
\[ \sin\xi_1\sin\xi_2\sin\xi_3 \left[a\cos\varphi + b\sin\varphi + M'_{222}\cos^3\varphi + (M'_{322} + M'_{232} +
M'_{223})\cos^2\varphi\sin\varphi \right. \] \[ \left. + (M'_{233} + M'_{323} +
M'_{332})\cos\varphi\sin^2\varphi + M'_{333}\sin^3\varphi)\right] = 0.
\]
Putting the independent coefficients of this trigonometric polynomial to zero, we obtain the additional
relations
\begin{eqnarray} \label{eq0}
M'_{222} &=& -a, \nonumber\\
M'_{322} + M'_{232} + M'_{223} &=& -b, \nonumber\\
M'_{233} + M'_{323} + M'_{332} &=& -a, \nonumber\\
M'_{333} &=& -b.
\end{eqnarray}

Now consider the condition $M' \in (B^{\otimes 3})^o$. This condition implies that the matrix formed of the
elements $M'_{1\alpha\beta}$, $\alpha,\beta = 1,2,3$ has singular values not exceeding 1, otherwise we would
find vectors $y,z \in B$ such that $M'(f_1,y,z) > 1$. The singular values of this matrix are given by
$\sqrt{a^2+b^2+c_1^2},0$. Here the first value has double multiplicity.
In a similar fashion we can consider the matrices formed of
the elements $M'_{\alpha1\beta}$, $M'_{\alpha\beta1}$, which yields the relations
\begin{equation} \label{ineq1}
a^2+b^2+c_1^2 \leq 1, \ a^2+b^2+c_2^2 \leq 1, \ a^2+b^2+c_3^2 \leq 1.
\end{equation}
A similar argument leads to the conclusion that the vector formed of the elements $M'_{\alpha23}$, $\alpha =
1,2,3$ must have a norm not exceeding 1. Permuting the indices, we get the inequalities
\[ c_1^2+{M'}^2_{223}+{M'}^2_{323} \leq 1,\ c_1^2+{M'}^2_{232}+{M'}^2_{332} \leq 1,\ c_2^2+{M'}^2_{223}+{M'}^2_{233} \leq 1, \]
\begin{equation} \label{ineq2}
\ c_2^2+{M'}^2_{322}+{M'}^2_{332} \leq 1,\ c_3^2+{M'}^2_{232}+{M'}^2_{233} \leq 1,\ c_3^2+{M'}^2_{322}+{M'}^2_{323} \leq 1.
\end{equation}
It follows that
\[ ||M'||^2 = 4(a^2+b^2) + 2(c_1^2+c_2^2+c_3^2) + {M'}^2_{322} + {M'}^2_{232} + {M'}^2_{223} +
{M'}^2_{233} + {M'}^2_{323} + {M'}^2_{332} \leq a^2+b^2 + 6.
\]
Finally using the condition $||M'||^2 = 7$, we get $a^2+b^2 = 1$ and inequalities (\ref{ineq1}),
(\ref{ineq2}) must actually be equalities. Hence there exists $\zeta$ such that $a = \cos\zeta$, $b =
\sin\zeta$, $c_1 = c_2 = c_3 = 0$, $|M'_{322}| = |M'_{232}| = |M'_{223}|$, $|M'_{233}| = |M'_{323}| =
|M'_{332}|$, $|M'_{322}|^2 + |M'_{233}|^2 = 1$. Combining this with (\ref{eq0}), we obtain $|M'_{322}| =
|\sin\zeta|$, $|M'_{233}| = |\cos\zeta|$ and equations (\ref{M'cond}) are verified.
\end{proof}

{\theorem \label{PTP4form} Let $B \subset {\mathbb R}^3$ be the unit ball. Then $\max_{T' \in (B^{\otimes 4})^o} ||T'|| = \sqrt{21}$. For any $T' \in (B^{\otimes 4})^o$ such that $||T'|| = \sqrt{21}$
there exist orthogonal $3 \times 3$ matrices $U,V,W,X$ such that $(U \otimes V \otimes W \otimes X)(T')$ can be obtained from the tensor $T$ defined by (\ref{T}) by some permutation of indices. }

\begin{proof}
The tensor $T$ defined by (\ref{T}) has norm $\sqrt{21}$. Hence $\max_{T' \in (B^{\otimes 4})^o} ||T'|| \geq \sqrt{21}$.

Let now $T' \in (B^{\otimes 4})^o$. Define three 3rd order tensors $M^1,M^2,M^3$ on $\mathbb R^3$ by
\begin{equation} \label{split}
M^k_{\alpha\beta\gamma} = T'_{\alpha\beta\gamma k}, \qquad \alpha,\beta,\gamma,k = 1,2,3.
\end{equation}
Then for any $u = (u_1,u_2,u_3)^T \in B$ we have $\sum_{k=1}^3 u_k M^k \in (B^{\otimes 3})^o$. In particular, $M^k \in (B^{\otimes 3})^o$ for all $k = 1,2,3$ and by Lemma \ref{PTP3form}
$||M^k||^2 \leq 7$. Therefore $||T'||^2 = \sum_{k=1}^3 ||M^k||^2 \leq 21$, which proves the first assertion of the lemma.

Suppose now in addition that $||T'|| = \sqrt{21}$. Then necessarily $||M^k||^2 = 7$ for all $k = 1,2,3$.
We will bring $T'$ to the form (\ref{T}) by successive application of orthogonal transformations in the factor spaces and permutations of the indices.

By Lemma \ref{PTP3form} there exist orthogonal transformations $U,V,W$ such that $(U \otimes V \otimes W)(M^1) = M$, where $M$ is given by (\ref{M}). Let us assume without restriction of generality that $M^1 = M$.
Let $x,y,z \in B$ be such that $M^1(x,y,z) = 1$. Then for any $u = (u_1,u_2,u_3)^T \in B$ we have $T'(x,y,z,u) = \sum_{k=1}^3 u_k M^k(x,y,z) = u_1 + u_2 M^2(x,y,z) + u_3 M^3(x,y,z) \leq 1$. Thus
the vector $(1,M^2(x,y,z),M^3(x,y,z))^T$ must have a norm not exceeding 1, and $\cos\varphi M^2(x,y,z) + \sin\varphi M^3(x,y,z) = 0$ for all $\varphi$.

Hence by Lemma \ref{nullcompl} for any $\varphi \in [-\pi,\pi]$ there exist $\zeta(\varphi) \in {\mathbb R}$, $\sigma_k(\varphi),\pi_k(\varphi) \in \{+1,-1\}$, $k = 1,2,3$ such that the tensor
$M' = \cos\varphi M^2 + \sin\varphi M^3$ satisfies conditions (\ref{M'cond}) with $\zeta = \zeta(\varphi)$, $\sigma_k = \sigma_k(\varphi)$, $\pi_k = \pi_k(\varphi)$, $k = 1,2,3$. In particular, we have
\[ \left( \begin{array}{cc} T'_{1122} & T'_{1123} \\ T'_{1132} & T'_{1133} \end{array} \right) \left( \begin{array}{c} \cos\varphi \\ \sin\varphi \end{array} \right) = \left( \begin{array}{c} \cos\zeta(\varphi) \\
\sin\zeta(\varphi) \end{array} \right) \qquad \forall\ \varphi.
\]
Hence the linear map given by above $2 \times 2$ coefficient matrix is norm preserving and consequently
orthogonal. Let us denote this coefficient matrix by $C$. Applying the orthogonal map $\diag(1,C)$ to the
last factor space, i.e.\ performing the transformation $T' \mapsto (I_3 \otimes I_3 \otimes I_3 \otimes
\diag(1,C))T'$, we make this coefficient matrix equal to the identity matrix $I_2$.

We can hence assume without restriction of generality that $T'_{1122} =  T'_{1133} = 1$, $T'_{1123} = T'_{1132} = 0$, and $\zeta(\varphi) = \varphi$.
Further we have $\cos\varphi T'_{3222} + \sin\varphi T'_{3223} = \sigma_1(\varphi) \sin\zeta(\varphi) = \sigma_1(\varphi) \sin\varphi$ for all $\varphi$. It follows that $\sigma_1(\varphi)$ is constant
as a function of $\varphi$. Similarly we can deduce that $\sigma_k(\varphi),\pi_k(\varphi)$ are constant for all $k = 1,2,3$. In the sequel we omit the dependence on $\varphi$ and write just $\sigma_k,\pi_k$.

In (\ref{split}) we defined the 3rd order tensors $M^k$ by putting the last index in $T'$ to $k$. We can also choose any of the other three indices and follow similar lines of reasoning. This leads
in addition to the relations $\sigma_1+\sigma_2+\sigma_3 = -1$, $\pi_1+\pi_2+\pi_3 = -1$ to the conditions $\pi_1+\sigma_2+\sigma_3 = -1$, $\sigma_1+\pi_2+\pi_3 = -1$, $\sigma_1+\pi_2+\sigma_3 = -1$,
$\pi_1+\sigma_2+\pi_3 = -1$, $\sigma_1+\sigma_2+\pi_3 = -1$, $\pi_1+\pi_2+\sigma_3 = -1$. Therefore $\sigma_k = \pi_k$ for all $k = 1,2,3$, and two of the numbers $\sigma_k$ equal $-1$, while the third one equals $1$.
By a permutation of the first three indices of $T'$ we can achieve that $\sigma_1 = 1$, $\sigma_2 = \sigma_3 = -1$.
Finally, if we now apply to $T'$ the map $\diag(1,-1,-1) \otimes I_3 \otimes I_3 \otimes I_3$, followed by a permutation of the first two indices,
we obtain the tensor $T$ defined by (\ref{T}).
\end{proof}

Lemma \ref{polarrad} now yields the following result.

{\corollary \label{PTP4formD} Let $B \subset {\mathbb R}^3$ be the unit ball. The largest ball that fits into the body $B^{\otimes 4}$ has radius $\sqrt{1/21}$. Let $Y \in \partial B^{\otimes 4}$ be of norm $\sqrt{1/21}$.
Then there exist orthogonal $3 \times 3$ matrices $U,V,W,X$ such that $21(U \otimes V \otimes W \otimes X)(Y)$ can be obtained from the tensor $T$ defined by (\ref{T}) by some permutation of indices.
\qed }

\section{Example of a mixed state of 4 qubits}
In this section we explicitly construct a mixed state of 4 qubits which has a distance of $\sqrt{1/336}$ from the maximally mixed state and
lies on the boundary to entanglement.

\medskip

For $m = 4$ the mixed state constructed in Corollary \ref{state_construction} is given by the matrix
\[
\rho = \frac{1}{168} \left( \begin{array}{cccccccccccccccc}
    11  &  0  &  0  &  0  &  0  &  0  &  0  &  0  &  0  &  0  &  0  &  0  &  0  &  0  &  0  &  0  \\
     0  & 10  &  1  &  0  & -1  &  0  &  0  &  0  &  1  &  0  &  0  &  0  &  0  &  0  &  0  &  0  \\
     0  &  1  & 10  &  0  &  1  &  0  &  0  &  0  & -1  &  0  &  0  &  0  &  0  &  0  &  0  &  0  \\
     0  &  0  &  0  & 11  &  0  & -1  &  1  &  0  &  0  &  1  & -1  &  0  &  0  &  0  &  0  &  0  \\
     0  & -1  &  1  &  0  & 10  &  0  &  0  &  0  &  1  &  0  &  0  &  0  &  0  &  0  &  0  &  0  \\
     0  &  0  &  0  & -1  &  0  & 11  & -1  &  0  &  0  & -1  &  4  &  0  & -1  &  0  &  0  &  0  \\
     0  &  0  &  0  &  1  &  0  & -1  & 11  &  0  &  0  &  0  & -1  &  0  &  1  &  0  &  0  &  0  \\
     0  &  0  &  0  &  0  &  0  &  0  &  0  & 10  &  0  &  0  &  0  &  1  &  0  & -1  &  1  &  0  \\
     0  &  1  & -1  &  0  &  1  &  0  &  0  &  0  & 10  &  0  &  0  &  0  &  0  &  0  &  0  &  0  \\
     0  &  0  &  0  &  1  &  0  & -1  &  0  &  0  &  0  & 11  & -1  &  0  &  1  &  0  &  0  &  0  \\
     0  &  0  &  0  & -1  &  0  &  4  & -1  &  0  &  0  & -1  & 11  &  0  & -1  &  0  &  0  &  0  \\
     0  &  0  &  0  &  0  &  0  &  0  &  0  &  1  &  0  &  0  &  0  & 10  &  0  &  1  & -1  &  0  \\
     0  &  0  &  0  &  0  &  0  & -1  &  1  &  0  &  0  &  1  & -1  &  0  & 11  &  0  &  0  &  0  \\
     0  &  0  &  0  &  0  &  0  &  0  &  0  & -1  &  0  &  0  &  0  &  1  &  0  & 10  &  1  &  0  \\
     0  &  0  &  0  &  0  &  0  &  0  &  0  &  1  &  0  &  0  &  0  & -1  &  0  &  1  & 10  &  0  \\
     0  &  0  &  0  &  0  &  0  &  0  &  0  &  0  &  0  &  0  &  0  &  0  &  0  &  0  &  0  & 11
\end{array} \right).
\]

{\corollary The matrix $\rho$ defined above describes a mixed 4-qubit state on the boundary to entanglement.
Its multiple $\frac{168}{11}\rho$ has a distance of $\sqrt{8/11}$ from the identity matrix $I_{16}$. }

\begin{proof}
The first statement follows from Corollary \ref{state_construction}. The second statement is an easily verifiable consequence
of similarity relations.
\end{proof}

This is the first explicit example of a mixed state on the boundary to entanglement whose multiple has a distance strictly
less than 1 from the identity matrix.

\end{document}